\documentstyle[12pt]{article}
\begin{document}

\title{\Large\bf BFFT quantization with nonlinear constraints} 
\author{J. Barcelos-Neto\thanks{\noindent e-mail:
barcelos@if.ufrj.br}\\ 
Instituto de F\'{\i}sica\\ 
Universidade Federal do Rio de Janeiro\\ 
RJ 21945-970 - Caixa Postal 68528 - Brasil\\
\\
\\
\date{}}

\maketitle
\abstract
We consider the method due to Batalin, Fradkin, Fradkina, and Tyutin
(BFFT) that makes the conversion of second-class constraints into
first-class ones for the case of nonlinear theories. We first present
a general analysis of an attempt to simplify the method, showing the
conditions that must be fulfilled in order to have first-class
constraints for nonlinear theories but that are linear in the
auxiliary variables.  There are cases where this simplification
cannot be done and the full BFFT method has to be used. However, in
the way the method is formulated, we show with details that it is not
practicable to be done. Finally, we speculate on a solution for these
problems.

\vfill
\vspace{1cm}
\newpage

\section{Introduction}
\renewcommand{\theequation}{1.\arabic{equation}}
\setcounter{equation}{0}
\bigskip

The quantization method due to Batalin, Fradkin, Fradkina, and
Tyutin~\cite{BF,BT} has as main purpose the conversion of
second-class constraints into first-class ones. This is achieved with
the aid of auxiliary variables that extend the phase space in a
convenient way.  After that, we have a gauge invariant system which
matches the original theory when the so-called unitary gauge is
chosen.

\medskip
The BFFT method is quite elegant and operates systematically. The
obtainment of first-class constraints is done in an iterative
process.  The first correction of the constraints is linear in the
auxiliary variables, the second one is quadratic, and so on. In the
case of systems with just linear constraints, we obtain that just
linear corrections are enough to make them first-class. Here, we
mention that the method is equivalent to express the dynamic
quantities by means of shifted coordinates~\cite{Das}.

\medskip
However, for systems with nonlinear constraints, the iterative
process may be go beyond the first correction. At this point resides
the problem we would like to address. The question is that the first
step of the process does not precisely fix the solution that shall be
used in the next steps. In recent papers, Banerjee et al. \cite{Ban}
have shown that an intelligent choice among the solutions displayed
at the first stage for the $O(N)$ nonlinear sigma-model and the
$CP^{N-1}$ can stop the iterative process there, avoiding the
possible inconveniences that could be found into the next steps.
This is a kind of simplification for the use of the method, but
unfortunately it cannot be applied for all cases.

\medskip
In the first part of this paper, we do a general analysis of the
conditions that makes possible the obtainment of linear first-class
constraints (in the auxiliary variables) for nonlinear theories. We
then show why the choice made by Banerjee et al. has worked on. We
concentrate on the $O(N)$ nonlinear sigma model (the $CP^{N-1}$ can
be verified in a similar way). After that, we deal with the
supersymmetric version of the $O(N)$ nonlinear sigma model and show
that it is also possible to have a linear solution in the auxiliary
variables for this case too, but being nonlocal. We deserve Sec. 3
for these developments and use Sec. 2 for a brief review of the BFFT
formalism in order to become clear the problem we would like to
address our attention. In Sec. 4, we consider some cases where these
choices are not or cannot be done and it is necessary to go to the
next order of the iterative process.  However, in a example we shall
discuss, there occurs an additional problem, besides the one related
to the choice of the solution that shall be used in the next steps:
we shall see that nor all solutions of the first process can lead to
a solution in the second one. If fact, we were just able to obtain a
very particular solution of the first step that could lead to a
solution in the second step. This fact show that the method is not
feasible to be used in the general case with nonlinear constraints.
We left an Appendix to contain some details of calculation and Sec. 5
for some concluding remarks.

\vspace{1cm}
\section{Brief review of the BFFT formalism}
\renewcommand{\theequation}{2.\arabic{equation}}
\setcounter{equation}{0}
\bigskip

Let us consider a system described by a Hamiltonian $H_0$ in a
phase-space $(q^i,p_i)$ with $i=1,\dots,N$. Here we suppose that the
coordinates are bosonic (extension to include fermionic degrees of
freedom and to the continuous case can be done in a straightforward
way). Let us also suppose that there just exist second-class
constraints. Denoting them by $T_a$, with $a=1,\dots M<2N$, we have

\begin{equation}
\label{2.1}
\bigl\{T_a,\,T_b\bigr\}=\Delta_{ab}\,,
\end{equation}

\bigskip\noindent
where $\det(\Delta_{ab})\not=0$. 

\medskip
As it was said, the general purpose of the BFFT formalism is to
convert second into first-class constraints. This is achieved by
introducing auxiliary canonical variables, one for each second-class
constraint (the connection between the number of second-class
constraints and the new variables in one-to-one is to keep the same
number of the physical degrees of freedom in the resulting extended
theory). We denote these auxiliary variables by $\eta^a$ and assume
that they have the following general structure

\begin{equation}
\label{2.2}
\bigl\{\eta^a,\,\eta^b\bigr\}=\omega^{ab}\,,
\end{equation}

\bigskip\noindent 
where $\omega^{ab}$ is a constant quantity with
$\det\,(\omega^{ab})\not=0$. The obtainment of $\omega^{ab}$ is
discussed in what follows. It is embodied in the calculation of the
resulting first-class constraints that we denote by $\tilde T_a$. Of
course, these depend on the new variables $\eta^a$, namely

\begin{equation}
\label{2.3}
\tilde T_a=\tilde T_a(q,p;\eta)\,,
\end{equation}

\bigskip\noindent 
and satisfy the boundary condition

\begin{equation}
\label{2.4}
\tilde T_a(q,p;0)=T_a(q,p)\,.
\end{equation}

\bigskip\noindent
Another characteristic of these new constraints is that they are
assumed to be strongly involutive, i.e.

\begin{equation}
\label{2.5}
\bigl\{\tilde T_a,\,\tilde T_b\bigr\}=0\,.
\end{equation}

\bigskip\noindent
We emphasize that this is a characteristic of the BFFT formalism,
i.e., the obtained first-class constraints are always supposed to
satisfy an Abelian algebra.

\medskip
The solution of~(\ref{2.5}) can be achieved by considering $\tilde T_a$
expanded as

\begin{equation}
\label{2.6}
\tilde T_a=\sum_{n=0}^\infty T_a^{(n)}\,,
\end{equation}

\bigskip\noindent
where $T_a^{(n)}$ is a term of order $n$ in $\eta$. Compatibility
with the boundary condition~(\ref{2.4}) requires

\begin{equation}
\label{2.7}
T_a^{(0)}=T_a\,.
\end{equation}

\bigskip\noindent
The replacement of~(\ref{2.6}) into~(\ref{2.5}) leads to a set of
equations, one for each coefficient of powers of $\eta$. We list some
of them below

\begin{eqnarray}
\bigl\{T_a^{(0)},\,T_b^{(0)}\bigr\}_{(q,p)}
&\!\!\!+\!\!\!&\bigl\{T_a^{(1)},\,T_b^{(1)}\bigr\}_{(\eta)}=0\,,
\label{2.8a}\\
\bigl\{T_a^{(0)},\,T_b^{(1)}\bigr\}_{(q,p)}
&\!\!\!+\!\!\!&\bigl\{T_a^{(1)},\,T_b^{(0)}\bigr\}_{(q,p)}
+\bigl\{T_a^{(1)},\,T_b^{(2)}\bigr\}_{(\eta)}
+\bigl\{T_a^{(2)},\,T_b^{(1)}\bigr\}_{(\eta)}=0,
\label{2.8b}\\
\bigl\{T_a^{(0)},\,T_b^{(2)}\bigr\}_{(q,p)}
&\!\!\!+\!\!\!&\bigl\{T_a^{(1)},\,T_b^{(1)}\bigr\}_{(q,p)}
+\bigl\{T_a^{(2)},\,T_b^{(0)}\bigr\}_{(q,p)}
+\bigl\{T_a^{(1)},\,T_b^{(3)}\bigr\}_{(\eta)}
\nonumber\\
&\!\!\!+\!\!\!&\bigl\{T_a^{(2)},\,T_b^{(2)}\bigr\}_{(\eta)}
+\bigl\{T_a^{(3)},\,T_b^{(1)}\bigr\}_{(\eta)}=0\,,
\label{2.8c}\\
&\vdots&
\nonumber
\end{eqnarray}

\bigskip\noindent 
The notation $\{,\}_{(q,p)}$ and $\{,\}_{(\eta)}$ represent the parts
of the Poisson bracket relative to the variables $(q,p)$ and
$(\eta)$, respectively.

\medskip
The BFFT method establishes that equations above are used iteratively
in the obtainment of the corrections $T^{(n)}$ ($n\geq1$).
Equation~(\ref{2.8a}) shall give us $T^{(1)}$. With this result
and~(\ref{2.8b}), one calculates $T^{(2)}$, and so on. To calculate
$T^{(1)}$ (that is linear in $\eta$), we may write

\begin{equation}
\label{2.9}
T_a^{(1)}=X_{ab}(q,p)\,\eta^b\,.
\end{equation}

\bigskip\noindent 
Introducing this expression into~(\ref{2.8a}) and using the boundary
condition~(\ref{2.4}), as well as~(\ref{2.1}) and~(\ref{2.2}), we get

\begin{equation}
\label{2.10}
\Delta_{ab}+X_{ac}\,\omega^{cd}\,X_{bd}=0\,.
\end{equation}

\bigskip\noindent 
We notice that this equation does not give $X_{ab}$ univocaly,
because it also contains the still unknown $\omega^{ab}$. What we
usually do is to choose $\omega^{ab}$ in such a way that the new
variables are unconstrained. It might be opportune to mention that
sometimes it is not possible to make a choice like that~\cite{Am}. In
this case, the new variables are constrained. In consequence, the
consistency of the method requires an introduction of other new
variables in order to transform these constraints also into
first-class. This may lead to an endless process. But it is important
to emphasize that $\omega^{ab}$ can be fixed anyway.

\medskip
However, even when one fixes $\omega^{ab}$ it is still not possible
to obtain a univocaly solution for $X_{ab}$. Let us check this point.
Since we are only considering bosonic coordinates\footnote{This
problem also exists in the fermionic sector.}, $\Delta_{ab}$ and
$\omega^{ab}$ are antisymmetric quantities. So, expression
(\ref{2.10}) compactly represents $M(M-1)/2$ independent equations.
On the other hand, there is no prior symmetry involving $X_{ab}$ and
they consequently represent a set of $M^2$ independent quantities.

\medskip
In the case where $X_{ab}$ do not depend on $(q,p)$, it is easily
seen that $T_a+T_a^{(1)}$ is already strongly involutive for any
choice we make and we are succeed in obtaining $\tilde T_a$ . If this
is not so, the usual procedure is to introduce $T_a^{(1)}$ into
equation~(\ref{2.8b}) to calculate $T_a^{(2)}$, and so on. At this
point resides the problem we had mentioned. We do not know a priori
what is the best choice we can make to go from one step to another.
More than that, we are going to show an example where solutions of
the first step do not necessary lead to a solution in the second one.

\medskip
What Banerjee et al. have done is a kind of simplification of the
BFFT method that consists to look for a convenient set of
coefficients, among the options contained in eq. (\ref{2.10}), in
such a way that the first-class constraints algebra could be obtained
in the first stage of the process. They were succeeded in the
particular case of the $O(N)$ nonlinear sigma model and $CP^{N-1}$
without Lagrange multiplier fields. In order to see in a general way
for what conditions this choice can be done, we the impose that the
linear constraints (in the auxiliary variables)

\begin{equation}
\tilde T_a=T_a+X_{ab}\,\eta^b
\label{2.11}
\end{equation}

\bigskip\noindent
satisfy the strong (Abelian) algebra (\ref{2.5}). Besides eq.
(\ref{2.10}), there remains the following equations to be satisfied

\begin{eqnarray}
&&\bigl\{T_a,X_{bc}\bigr\}+\bigl\{X_{ac},T_b\bigr\}=0\,,
\label{2.12}\\
&&\bigl\{X_{ac},X_{bd}\bigr\}+\bigl\{X_{ad},X_{bc}\bigr\}=0\,.
\label{2.13}
\end{eqnarray}

\bigskip\noindent
We notice that when $X_{ab}$ do not depend on the phase space
coordinates, the equations above are satisfied trivially. However,
when this is not so, we have more equations than variables and each
case has to be verified in a separate way. We are going to discuss in
the next section why the $O(N)$ nonlinear sigma model can be
considered as a particular example where eqs. (\ref{2.12}) and
(\ref{2.13}) are verified.

\vspace{1cm}
\section{Simplified use of the BFFT formalism}
\renewcommand{\theequation}{3.\arabic{equation}}
\setcounter{equation}{0}
\bigskip

Let us concentrate here in the $O(N)$ nonlinear sigma-model. It is
described by the Lagrangian

\begin{equation}
{\cal L}=\frac{1}{2}\,\partial_\mu\phi^A\partial^\mu\phi^A
+\frac{1}{2}\,\lambda\,\bigl(\phi^A\phi^A-1\bigr)\,,
\label{3.1}
\end{equation}

\bigskip\noindent
where the index $A$ is related to the $O(N)$ symmetry group. Let us
obtain the constraints~\cite{Dirac}. The canonical momenta are 

\begin{eqnarray}
\pi^A&=&\frac{\partial{\cal L}}{\partial\dot\phi^A}=\dot\phi^A\,,
\label{3.2}\\
p&=&\frac{\partial{\cal L}}{\partial\dot\lambda}=0\,.
\label{3.3}
\end{eqnarray}

\bigskip\noindent
We notice that (\ref{3.3}) is a primary constraint. In order to look
for secondary constraints we construct the Hamiltonian

\begin{eqnarray}
{\cal H}&=&\pi^A\dot\phi^A+p\dot\lambda-{\cal L}+\xi p\,,
\nonumber\\
&=&\frac{1}{2}\,\pi^A\pi^A
+\frac{1}{2}\phi^{A\prime}\phi^{A\prime}
-\frac{1}{2}\,\lambda\,\bigl(\phi^A\phi^A-1\bigr)
+\tilde\xi p\,,
\label{3.4}
\end{eqnarray}

\bigskip\noindent
where the velocity $\dot\lambda$ was absorbed in the Lagrange
multiplier $\tilde\xi$ by the redefinition
$\tilde\xi=\xi+\dot\lambda$. The consistency condition for the
constraint $p=0$ leads to another constraint

\begin{equation}
\phi^A\phi^A-1=0\,.
\label{3.5}
\end{equation}

\bigskip\noindent
At this stage, we have two options. The first one is to introduce the
constraint above into the Hamiltonian by means of another Lagrange
multiplier. The result is 

\begin{eqnarray}
{\cal H}&=&\frac{1}{2}\,\pi^A\pi^A
+\frac{1}{2}\,\phi^{A\prime}\phi^{A\prime}
-\frac{1}{2}\,\lambda\,\bigl(\phi^A\phi^A-1\bigr)
+\tilde\xi\,p
+\zeta\,\bigl(\phi^A\phi^A-1\bigr)\,,
\nonumber\\
&=&\frac{1}{2}\,\pi^A\pi^A
+\frac{1}{2}\,\phi^{A\prime}\phi^{A\prime}
+\tilde\xi\,p
+\tilde\zeta\,\bigl(\phi^A\phi^A-1\bigr)\,.
\label{3.6}
\end{eqnarray}

\bigskip\noindent

The field $\lambda$ was also absorbed by the Lagrange multiplier
$\zeta$ ($\tilde\zeta=\zeta-\frac{1}{2}\lambda$). The consistency
condition for the constraint (\ref{3.5}) leads to another more
constraint 

\begin{equation}
\phi^A\pi^A=0\,.
\label{3.7}
\end{equation}

\bigskip\noindent
We mention that the consistency condition for this constraint will
give us the Lagrange multiplier $\tilde\zeta$ and no more
constraints.  Since we have absorbed the field $\lambda$, its
momentum $p$ does not play any role in the theory and we can
disregard it by using the constraint relation (\ref{3.3}) in a strong
way. So the constraints of the theory is this case are

\begin{eqnarray}
T_1&=&\phi^A\phi^A-1\,,
\nonumber\\
T_2&=&\pi^A\phi^A\,.
\label{3.8}
\end{eqnarray}

\bigskip
The other option we could have followed was to keep the Lagrange
multiplier $\lambda$ in the theory. To do this, we consider that the
constraint (\ref{3.5}) is already in the Hamiltonian due to the
presence of the term $-\,\frac{\lambda}{2}\,(\phi^A\phi^A-1)$. So,
instead of the Hamiltonian (\ref{3.6}) we use the previous one given
by (\ref{3.4}) in order to verify the consistency condition  of the
constraint (\ref{3.5}). It is easily seen that the constraint
(\ref{3.7}) is obtained again, and the consistency condition for
it leads to a new constraint 

\begin{equation}
\pi^A\pi^A+\phi^A\phi^{A\prime\prime}
+\lambda\,\phi^A\phi^A=0\,.
\label{3.9}
\end{equation}

\bigskip
No more constraints are obtained. So, according to the second option,
the constraints of the theory are

\begin{eqnarray}
T_1&=&\phi^A\phi^A-1\,,
\nonumber\\
T_2&=&\pi^A\phi^A\,,
\nonumber\\
T_3&=&p\,,
\nonumber\\
T_4&=&\pi^A\pi^A+\lambda\,\phi^A\phi^A
+\phi^A\phi^{A\prime\prime}\,.
\label{3.10}
\end{eqnarray}

\bigskip
We mention that the set given by (\ref{3.10}) is an example where it
is necessary to go to higher order of the iterative process of the
BFFT method. We shall discuss this example in the next section. Let
us concentrate here on the first set of constraints (\ref{3.8}). The
quantity $\Delta_{ab}$, defined by eq. (\ref{2.1}), can be directly
calculated. The result is

\begin{equation}
\Delta_{ab}=2\,\left(\begin{array}{cc}
0&1\\
-1&0\\
\end{array}\right)\,
\phi^A\phi^A\,\delta(x-y)\,.
\label{3.11}
\end{equation}

\bigskip\noindent
At this point we could be tempted to use back the constraint $T_1$
in order to simplify the expression above. This cannot be done
because expressions (\ref{2.8a})-(\ref{2.8c}) have to be verified in
a strong way.

\medskip
Let us extend the phase space by introducing two new variables
$(\eta^1,\,\eta^2)$ and consider that $\eta^2$ is the canonical
momentum conjugate to $\eta^1$. Consequently, expression (\ref{2.10})
gives us just one equation

\begin{equation}
X_{12}X_{21}-X_{11}X_{22}=2\,\phi^A\phi^A\,\delta(x-y)\,,
\label{3.12}
\end{equation}

\bigskip\noindent
where there are four quantities to be fixed. The simplified used of
the BFFT formalism requires that these quantities also satisfy
expressions (\ref{2.12}) and (\ref{2.13}), what correspond to five
more equations.

\medskip
Eventhough there are more equations (six) than variables (four), the
system presents solution, as has been pointed out by Banerjee et al.
but considering a different analysis. The arguments we use are the
following.  Looking at eq. (\ref{3.12}), we notice that at least one
of the variables must contain the field $\phi^A$. This is not a
problem to satisfy expressions (\ref{2.13}). However, concerning the
ones given (\ref{2.12}) we need some care because the constraint
$T_2$ contains the momentum $\pi^A$. The way we have to satisfy
expression (\ref{2.12}) for $a=2$ (the index where $T_2$ will appear)
is to use the fact that (\ref{2.12}) is antisymmetric in the indices
$a$ and $b$. So, if we choose the coefficients $X_{2c}$ to contain
the fields $\phi^A$, equations (\ref{2.12}) will be automatically
verified. The Banerjee et al. choice was (up to some multiplicative
factors)

\begin{eqnarray}
X_{11}&=&2\,\delta(x-y)\,,
\nonumber\\
X_{22}&=&-\,\phi^A\phi^A\,\delta(x-y)\,,
\nonumber\\
X_{12}&=&X_{21}=0\,,
\label{3.13}
\end{eqnarray}

\bigskip\noindent
that leads to the following set of first-class constraints 

\begin{eqnarray}
\tilde T_1&=&\phi^A\phi^A-1+2\,\eta^1\,,
\nonumber\\
\tilde T_2&=&\phi^A\pi^A-\phi^A\phi^A\,\eta^2\,.
\label{3.14}
\end{eqnarray}

\bigskip\noindent
Another general choice could be to use the coefficients $X_{21}$ to
contain $\phi^A$, that is

\begin{eqnarray}
X_{12}&=&2\,\delta(x-y)\,,
\nonumber\\
X_{21}&=&\phi^A\phi^A\,\delta(x-y)\,,
\nonumber\\
X_{11}&=&X_{22}=0\,.
\label{3.15}
\end{eqnarray}

\bigskip\noindent
The second choice is nothing other than the interchange of $\eta^1$
and $\eta^2$ in equations (\ref{3.14}), accompanied by a convenient
change of signs. In any case, both solutions close in a Abelian
algebra like (\ref{2.5}).

\bigskip
An interesting point is that the simplified used of the BFFT
formalism can also be applied to the supersymmetric case. The
constraints in this case are (still without Lagrange multipliers)

\begin{eqnarray}
T_1&=&\phi^A\phi^A-1\,,
\nonumber\\
T_2&=&\phi^A\pi^A\,,
\nonumber\\
T_{3\alpha}&=&\phi^A\psi^A_\alpha\,,
\label{3.16}
\end{eqnarray}

\bigskip\noindent
where the canonical momentum conjugate to $\psi^A_\alpha$ is a
constraint relation that can be eliminated by using the Dirac bracket
\footnote{We could have used auxiliary variables to convert
these constraints into first-class too. But this would be a trivial
operation that would not lead to any new significant result.}
\cite{Dirac}

\begin{equation}
\bigl\{\psi_\alpha^A(x),\,\psi_\beta^B(y)\bigr\}
=-\,i\,\delta_{\alpha\beta}\,\delta^{AB}\delta(x-y)\,.
\label{3.17}
\end{equation}

\bigskip\noindent
We thus have

\begin{eqnarray}
\Delta_{12}&=&2\,\phi^A\phi^A\,\delta(x-y)\,,
\nonumber\\
\Delta_{23}&=&-\,\phi^A\psi^A_\alpha\,\delta(x-y)\,,
\nonumber\\
\Delta_{3\alpha\,3\beta}&=&-\,i\,\delta_{\alpha\beta}\,
\phi^A\phi^A\,\delta(x-y)\,.
\label{3.18}
\end{eqnarray}

\bigskip\noindent
The remaining quantities vanish. We extend the phase-space by
introducing the coordinates: $\eta_1$, $\eta_2$ and $\chi_\alpha$ and
consider that they satisfy the fundamental relations

\begin{eqnarray}
\bigl\{\eta_1,\,\eta_2\bigr\}&=&\delta(x-y)\,,
\nonumber\\
\bigl\{\chi_\alpha,\,\chi_\beta\bigr\}&=&-\,i\,
\delta_{\alpha\beta}\,\delta(x-y)\,.
\label{3.19}
\end{eqnarray}

\bigskip
Let us verify if it is also possible to have a solution for equations
(\ref{2.10}), (\ref{2.12}) and (\ref{2.13}). These have to be
slightly modified by virtue of the presence of fermionic constraints,
namely

\begin{eqnarray}
&&\Delta_{ab}+(-1)^{(\epsilon_b+1)\epsilon_d}\,
X_{ac}\,\omega^{cd}\,X_{bd}=0\,,
\label{3.20}\\
&&\bigl\{T_a,X_{bc}\bigr\}+(-1)^{\epsilon_b\epsilon_c}\,
\bigl\{X_{ac},T_b\bigr\}=0\,,
\label{3.21}\\
&&\bigl\{X_{ac},X_{bd}\bigr\}+\bigl\{X_{ad},X_{bc}\bigr\}=0\,,
\label{3.22}
\end{eqnarray}

\bigskip\noindent
where $\epsilon_a=0$ for $a=1,2$ (bosonic constraints) and
$\epsilon_a=1$ otherwise (fermionic constraints). The expression
(\ref{3.20}) yields the following set of equations

\begin{eqnarray}
X_{12}X_{21}-X_{11}X_{22}-iX_{1\,3\alpha}X_{2\,3\alpha}
&=&2\,\phi^A\phi^A\,\delta(x-y)\,,
\label{3.23}\\
X_{21}X_{3\alpha\,2}-X_{22}X_{3\alpha\,1}
-iX_{2\,3\beta}X_{3\alpha\,3\beta}
&=&\phi^A\psi^A_\alpha\,\delta(x-y)\,,
\label{3.24}\\
X_{3\alpha\,1}X_{3\beta\,2}-X_{3\alpha\,2}X_{3\beta\,1}
-iX_{3\alpha\,3\gamma}X_{3\beta\,3\gamma}
&=&i\,\delta_{\alpha\beta}\,\phi^A\phi^A\,\delta(x-y)\,,
\label{3.25}\\
X_{11}X_{3\alpha\,2}-X_{12}X_{3\alpha\,1}
-iX_{1\,3\beta}X_{3\alpha\,3\beta}
&=&0\,.
\label{3.26}
\end{eqnarray}

\bigskip\noindent
From equation (\ref{3.25}) we are forced to conclude that

\begin{equation}
X_{3\alpha\,3\beta}^A=i\,\delta_{\alpha\beta}
\,\sqrt{\phi^A\phi^A}\,\delta(x-y)\,,
\label{3.27}
\end{equation}

\bigskip\noindent
and a careful analysis of the remaining equations permit us to
infer that a solution that is also compatible with (\ref{3.21}) and
(\ref{3.22}) is given by

\begin{eqnarray}
X_{11}&=&2\,\delta(x-y)\,,
\label{3.28}\\
X_{22}&=&-\,\phi^A\phi^A\,\delta(x-y)\,,
\label{3.29}\\
X_{2\,3\alpha}&=&\frac{\phi^A\psi^A_\alpha}
{\sqrt{\phi^B\phi^B}}\,\delta(x-y)\,.
\label{3.30}
\end{eqnarray}

\bigskip\noindent
The remaining coefficients are zero. The choice given by equations
(\ref{3.27})-(\ref{3.30}) leads to the following set of first-class
constraints 

\begin{eqnarray}
\tilde T_1&=&\phi^A\phi^A-1+2\eta^1\,,
\nonumber\\
\tilde T_2&=&\phi^A\pi^A-\phi^A\phi^A\,\eta^2
+\frac{\phi^A\psi^A_\alpha\chi_\alpha}
{\sqrt{\phi^B\phi^B}}
\nonumber\\
\tilde T_{3\alpha}&=&\phi^A\psi^A_\alpha
+i\,\sqrt{\phi^A\phi^A}\,\chi_\alpha\,.
\label{3.31}
\end{eqnarray}

\bigskip\noindent
The above result shows us that it is also possible to obtain a
solution, eventhough nonlocal, for the simplified version of the BFFT
method when applied to the supersymmetric nonlinear sigma model.

\vspace{1cm}
\section{Beyond the first interative step}
\renewcommand{\theequation}{4.\arabic{equation}}
\setcounter{equation}{0}

\bigskip
In order to have a better understanding of the problem we are going
to deal with, let us first consider the case without Lagrange
multipliers, and take a different choice of the one given by
Banerjee et al. For example, let us suppose that instead of solutions
(\ref{3.13}) or (\ref{3.15}), we take

\begin{eqnarray}
X_{11}&=&\phi^A\phi^A\,\delta(x-y)\,,
\nonumber\\
X_{22}&=&-\,2\,\delta(x-y)\,,
\nonumber\\
X_{12}&=&X_{21}\,\,=\,\,0\,.
\label{4.1}
\end{eqnarray}

\bigskip\noindent
For this choice, we have

\begin{eqnarray}
T_1^{(1)}&=&\phi^A\phi^A\,\eta^1\,,
\nonumber\\
T_2^{(1)}&=&-\,2\eta^2\,.
\label{4.2}
\end{eqnarray}

\bigskip\noindent
We easily see that eq. (\ref{2.12}) is not more verified and
consequently the quantities $T_1+T_1^{(1)}$ and $T_2+T_2^{(1)}$ do
not form first-class constraints anymore. It is then necessary to go
to the second step of the BFFT method, what corresponds to use the
equation (\ref{2.8b}), namely

\begin{equation}
\bigl\{T_a,\,T_b^{(1)}\bigr\}_{(\phi,\pi)}
+\bigl\{T_a^{(1)},\,T_b\bigr\}_{(\phi,\pi)}
+\bigl\{T_a^{(1)},\,T_b^{(2)}\bigr\}_{(\eta)}
+\bigl\{T_a^{(2)},\,T_b^{(1)}\bigr\}_{(\eta)}=0\,.
\label{4.3}
\end{equation}

\bigskip\noindent
The combination of expressions (\ref{3.8}), (\ref{4.2}), and
(\ref{4.3}) permit us to infer the following solution for $T_a^{(2)}$

\begin{eqnarray}
T_1^{(2)}&=&\eta^2\eta^2\,,
\nonumber\\
T_2^{(2)}&=&-\,2\,\eta^1\eta^2\,.
\label{4.4}
\end{eqnarray}

\bigskip\noindent
We then notice that the quantities $\tilde
T_a=T_a+T_a^{(1)}+T_a^{(2)}$, namely

\begin{eqnarray}
\tilde T_1&=&\phi^A\phi^A-1+\phi^A\phi^A\,\eta^1+\eta^2\eta^2\,,
\nonumber\\
\tilde T_2&=&\phi^A\pi^A-2\eta^2-2\eta^1\eta^2
\label{4.5}
\end{eqnarray}

\bigskip\noindent
are first-class constraints and close in an Abelian algebra. Thus,
without the choice made by Banerjee et al., it is necessary to go to
the second order of the iterative process. The obtained
first-class constraints, for this particular version of the $O(N)$
nonlinear sigma-model, are quadratic in the auxiliary variables.

\bigskip
The same analysis could also have been done for the supersymmetric
case (without Lagrange multipliers). We just mention that the set of
first-class constrains is given by

\begin{eqnarray}
\tilde T_1&=&\phi^A\phi^A\bigl(1+\eta^1\bigr)-1\,,
\nonumber\\
\tilde T_2&=&\phi^A\pi^A-2\,\eta^2-2\,\eta^1\eta^2
+\frac{\phi^A\psi^A_\alpha\chi_\alpha}
{\sqrt{\phi^B\phi^B}}\,,
\nonumber\\
\tilde T_{3\alpha}&=&\phi^A\psi^A_\alpha
+i\,\sqrt{\phi^A\phi^A}\,\chi_\alpha\,.
\label{4.6}
\end{eqnarray}

\bigskip
Let us now consider the more general version of the $O(N)$ nonlinear
sigma model, which corresponds to the set of constraints given by
expressions (\ref{3.10}). First we write down the quantities
$\Delta_{ab}$. The nonvanishing ones are

\begin{eqnarray}
\Delta_{12}(x,y)&=&2\,\phi^A\phi^A\,\delta(x-y)\,,
\nonumber\\
\Delta_{14}(x,y)&=&4\,\phi^A\pi^A\,\delta(x-y)\,,
\nonumber\\
\Delta_{24}(x,y)&=&2\,\bigl(\pi^A\pi^A-\lambda\,\phi^A\phi^A
-\phi^A\phi^{A\prime\prime}\bigr)\,\delta(x-y)
\nonumber\\
&&-\,2\,\phi^A\phi^{A\prime}\,\delta^\prime(x-y)
-\phi^A\phi^A\,\delta^{\prime\prime}(x-y)\,,
\nonumber\\
\Delta_{34}(x,y)&=&-\,\phi^A\phi^A\,\delta(x-y)\,,
\label{4.7}
\end{eqnarray}

\bigskip\noindent
where prime always means derivative with respect do $x$. Here, we
extend the phase-space by introducing four coordinates $\eta^a$ and
consider that $\eta^3$ and $\eta^4$ are the canonical momenta
conjugated to $\eta^1$ and $\eta^2$, respectively. Thus, the matrix
$(\omega^{ab})$ reads

\begin{equation}
\bigl(\omega^{ab}\bigr)
=\left(\begin{array}{cccc}
0&0&1&0\\
0&0&0&1\\
-1&0&0&0\\
0&-1&0&0\\
\end{array}\right)\,\delta(x-y)\,.
\label{4.8}
\end{equation}

\bigskip\noindent
The combination of (\ref{2.10}), (\ref{4.7}) and (\ref{4.8}) gives
us the set of equations

\begin{eqnarray}
X_{11}X_{23}+X_{12}X_{24}-X_{13}X_{21}-X_{14}X_{22}
&=&-\,2\,\phi^A\phi^A\,\delta\,,
\nonumber\\
X_{11}X_{33}+X_{12}X_{34}-X_{13}X_{31}-X_{14}X_{32}
&=&0\,,
\nonumber\\
X_{11}X_{43}+X_{12}X_{44}-X_{13}X_{41}-X_{14}X_{42}
&=&-\,4\,\phi^A\pi^A\,\delta\,,
\nonumber\\
X_{21}X_{33}+X_{22}X_{34}-X_{23}X_{31}-X_{24}X_{32}
&=&0\,,
\nonumber\\
X_{21}X_{43}+X_{22}X_{44}-X_{23}X_{41}-X_{24}X_{42}
&=&-\,2\,\bigl(\pi^A\pi^A-\lambda\phi^A\phi^A
-\phi^A\phi^{A\prime\prime}\bigr)\,\delta
\nonumber\\
&&+\,2\,\phi^A\phi^{A\prime}\,\delta^\prime
+\phi^A\phi^A\,\delta^{\prime\prime}
\nonumber\\
X_{31}X_{43}+X_{32}X_{44}-X_{33}X_{41}-X_{34}X_{42}
&=&\phi^A\phi^A\,\delta\,.
\label{4.9}
\end{eqnarray}

\bigskip\noindent
where $\delta$, $\delta^\prime$, $\delta^{\prime\prime}$ stand for
simplified notations of the delta function and its derivatives.  We
easily observe that the arguments used in the case without the
Lagrange multiplier fields cannot be applied here. Now, the
coefficients $X_{ab}$ contain momenta and it is not trivial that eq.
(\ref{2.13}) can be verified.  Further, we also have momenta in more
than one constraint and this consequently means that we cannot fix
one of eqs. (\ref{2.12}) without spoiling the others. This is an
example where it actually necessary to go to higher order of the
iterative process of the BFFT method.

\medskip
However, an interesting and, in some sense, unexpected fact occurs:
nor all solution obtained in the first stage leads to a solution in
the second one.  We shall discuss this point with details in the
Appendix. Let us just write down a convenient solution of the first
step of the method (that leads to a solution into the second step)

\begin{eqnarray}
&&X_{11}=-\,2\,\phi^A\phi^A\,\delta\,,
\hspace{.5cm}X_{12}=0\,,
\hspace{.5cm}X_{13}=0\,,
\hspace{.5cm}X_{14}=4\,\phi^A\pi^A\,\delta\,,
\nonumber\\
&&X_{21}=0\,,
\hspace{.5cm}X_{22}=0\,,
\hspace{.5cm}X_{23}=\delta\,,
\hspace{.5cm}X_{24}=2\,\pi^A\pi^A\,\delta\,,
\nonumber\\
&&X_{31}=0\,,
\hspace{.5cm}X_{32}=0\,,
\hspace{.5cm}X_{33}=0\,,
\hspace{.5cm}X_{34}=-\,\phi^A\phi^A\,\delta
\nonumber\\
&&X_{41}=-\,2\,\lambda\phi^A\phi^A\,\delta
-2\,\phi^A\phi^{A\prime\prime}\,\delta
-\,2\,\phi^A\phi^{A\prime}\,\delta^\prime
-\,\phi^A\phi^A\,\delta^{\prime\prime}\,,
\nonumber\\
&&X_{42}=\delta\,,
\hspace{.5cm}X_{43}=0\,,
\hspace{.5cm}X_{44}=-\,4\,\phi^{A\prime}\pi^A\,\delta^\prime\,,
\label{4.10}
\end{eqnarray}

\bigskip\noindent
which leads to the following first order correction of the
constraints 

\begin{eqnarray}
T_1^{(1)}&=&-\,2\,\phi^A\phi^A\,\eta^1+4\,\phi^A\pi^A\,\eta^4\,,
\nonumber\\
T_2^{(1)}&=&\eta^3+2\,\pi^A\pi^A\,\eta^4\,,
\nonumber\\
T_3^{(1)}&=&-\,\phi^A\phi^A\,\eta^4\,,
\nonumber\\
T_4^{(1)}&=&-\,2\,\lambda\,\phi^A\phi^A\,\eta^1
-\,2\,\phi^A\phi^{A\prime\prime}\,\eta^1
-\,2\,\phi^A\phi^{A\prime}\,\eta^{1\prime}\,.
\nonumber\\
&&-\,\phi^A\phi^A\,\eta^{1\prime\prime}+\eta^2
-\,4\,\phi^{A\prime}\pi^A\,\eta^{4\prime}
\label{4.11}
\end{eqnarray}

\bigskip\noindent
It is important to emphasize that the coefficients $X_{21}$ and
$X_{44}$ could have taken any value. For the former, we choose the
simplest possibility and make it zero. For the last one, this would
not work. If we had also taken it as zero we would run into
difficulties in the second step of the method. The choice we have
made was to get a term like
$8\,\phi^A\phi^{A\prime}\eta^{4\prime}\,\delta$ into eq.(\ref{4.3})
for $a=1$ and $b=4$. This and the term
$8\,\phi^A\phi^{A\prime}\eta^4\,\delta^\prime$, that appears in the
same equation and that could not be matched without spoiling what was
already fixed, can be put together as
$8\,\phi^A\phi^A\,(\eta^4\delta)^\prime$. This result makes possible
to fix $T_1^{(2)}$ and $T_4^{(2)}$. For details of a similar
calculation, see the example discussed in the Appendix.

\bigskip
The combination of (\ref{3.10}), (\ref{4.3}), and (\ref{4.11}) gives 

\begin{eqnarray}
&&\bigl\{-2\,\phi^A\phi^A\eta^1+4\,\phi^A\pi^A\eta^4,
T_2^{(2)}\bigr\}_{(\eta)}
+\bigl\{T_1^{(2)},\eta^3+2\,\pi^A\pi^A\eta^4\bigr\}_{(\eta)}
\nonumber\\
&&\phantom{\bigl\{-2\phi^A\phi^A}
=\,\,4\,\phi^A\phi^A\eta^1\,\delta-8\,\phi^A\pi^A\eta^4\,\delta\,,
\label{4.12}\\
&&\bigl\{-2\,\phi^A\phi^A\eta^1+4\,\phi^A\pi^A\eta^4,
T_3^{(2)}\bigr\}_{(\eta)}
+\bigl\{T_1^{(2)},-\phi^A\phi^A\eta^4\bigr\}_{(\eta)}=0\,,
\label{4.13}\\
&&\bigl\{-2\,\phi^A\phi^A\eta^1+4\,\phi^A\pi^A\eta^4,
T_4^{(2)}\bigr\}_{(\eta)}
+\bigl\{T_1^{(2)},-2\,\lambda\phi^A\phi^A\eta^1
-2\,\phi^A\phi^{A\prime\prime}\eta^1
\nonumber\\
&&\phantom{\bigl\{-2\phi^A\phi^A\eta^1}
-2\,\phi^A\phi^{A\prime}\eta^{1\prime}
-\phi^A\phi^A\eta^{1\prime\prime}
+\eta^2-4\,\phi^{A\prime}\pi^A\eta^{4\prime}\bigr\}_{(\eta)}
\nonumber\\
&&\phantom{\bigl\{-2\phi^A\phi^A}
=\,\,8\,\phi^A\pi^A\eta^1\,\delta
-8\,\pi^A\pi^A\eta^4\,\delta
+8\,\lambda\phi^A\phi^A\eta^4\,\delta
+8\,\phi^A\phi^{A\prime\prime}\eta^4\,\delta
\nonumber\\
&&\phantom{\bigl\{-2\phi^A\phi^A\eta^1}
+8\,\phi^A\phi^{A\prime}\eta^{4\prime}\,\delta
+8\,\phi^A\phi^{A\prime}\eta^4\,\delta^\prime
+4\,\phi^A\phi^A\eta^4\,\delta^{\prime\prime}\,,
\label{4.14}\\
&&\bigl\{\eta^3+2\,\pi^A\pi^A\eta^4,T_3^{(2)}\bigr\}_{(\eta)}
+\bigl\{T_2^{(2)},-\phi^A\phi^A\eta^4\bigr\}_{(\eta)}
=-2\,\phi^A\phi^A\eta^4\,\delta\,,
\label{4.15}\\
&&\bigl\{\eta^3+2\,\pi^A\pi^A\eta^4,T_4^{(2)}\bigr\}_{(\eta)}
+\bigl\{T_2^{(2)},-2\,\lambda\phi^A\phi^A\eta^1
-2\,\phi^A\phi^{A\prime\prime}\eta^1
\nonumber\\
&&\phantom{\bigl\{-2\phi^A\phi^A\eta^1}
-2\,\phi^A\phi^{A\prime}\eta^{1\prime}
-\phi^A\phi^A\eta^{1\prime\prime}
+\eta^2-4\,\phi^{A\prime}\pi^A\eta^{4\prime}\bigr\}_{(\eta)}
\nonumber\\
&&\phantom{\bigl\{-2\phi^A\phi^A}
=\,\,-\,4\,\lambda\phi^A\phi^A\eta^1\,\delta
-4\,\phi^A\phi^{A\prime\prime}\eta^1\,\delta
-4\,\phi^A\phi^{A\prime}\eta^{1\prime}\,\delta
\nonumber\\
&&\phantom{\bigl\{-2\phi^A\phi^A\eta^1}
-\,2\,\phi^A\phi^A\eta^{1\prime\prime}\,\delta
+8\,\lambda\phi^A\pi^A\eta^4\,\delta
+8\,\phi^{A\prime\prime}\pi^A\eta^4\,\delta
\nonumber\\
&&\phantom{\bigl\{-2\phi^A\phi^A\eta^1}
-\,4\,\phi^{A\prime}\pi^A\eta^{4\prime}\,\delta
-4\,\phi^A\pi^{A\prime}\eta^{4\prime}\,\delta
-4\,\phi^A\pi^A\eta^{4\prime\prime}\,\delta
\nonumber\\
&&\phantom{\bigl\{-2\phi^A\phi^A\eta^1}
-\,4\,\phi^A\phi^{A\prime}\eta^1\,\delta^\prime
-2\,\phi^A\phi^A\eta^{1\prime}\,\delta^\prime
+8\,\phi^{A\prime}\pi^A\eta^4\,\delta^\prime
\nonumber\\
&&\phantom{\bigl\{-2\phi^A\phi^A\eta^1}
-4\,\phi^A\pi^A\eta^{4\prime}\,\delta^\prime
-2\,\phi^A\phi^A\eta^1\,\delta^{\prime\prime}
+4\,\phi^A\pi^A\eta^4\,\delta^{\prime\prime}\,,
\label{4.16}\\
&&\bigl\{-\phi^A\phi^A\eta^4,T_4^{(2)}\bigr\}_{(\eta)}
+\bigl\{T_3^{(2)},-2\,\lambda\phi^A\phi^A\eta^1
-2\,\phi^A\phi^{A\prime\prime}\eta^1
\nonumber\\
&&\phantom{\bigl\{-2\phi^A\phi^A\eta^1}
-\,2\,\phi^A\phi^{A\prime}\eta^{1\prime}
-\phi^A\phi^A\eta^{1\prime\prime}
+\eta^2-4\phi^{A\prime}\pi^A\eta^{4\prime}\bigr\}_{(\eta)}
\nonumber\\
&&\phantom{\bigl\{-2\phi^A\phi^A}
=\,\,-2\,\phi^A\phi^A\eta^1\,\delta
+4\,\phi^A\pi^A\eta^4\,\delta
\label{4.17}\end{eqnarray}

\bigskip\noindent
After a hard and careful algebraic work, we obtain the following
solution for this set of equations

\begin{eqnarray}
T_1^{(2)}&=&-\,8\,\phi^A\pi^A\eta^1\eta^4
+4\,\bigl(\pi^A\pi^A-\phi^A\phi^{A\prime\prime}
-\lambda\,\phi^A\phi^A\bigr)\eta^4\eta^4
\nonumber\\
&&-\,8\,\phi^A\phi^{A\prime}\,\eta^4\eta^{4\prime}
-\,4\,\phi^A\phi^A\,\eta^4\eta^{4\prime\prime}
\nonumber\\
T_2^{(2)}&=&-\,2\,\eta^1\eta^3
+4\,\bigl(\phi^A\phi^{A\prime\prime}
+\phi^{A\prime}\phi^{A\prime}\bigr)\,\eta^1\eta^4
+4\,\phi^A\phi^{A\prime}\,\eta^{1\prime}\eta^4
\nonumber\\
&&+\,2\,\phi^A\phi^A\,\eta^{1\prime\prime}\eta^4
+\,8\,\phi^A\phi^{A\prime}\eta^1\eta^{4\prime}
+\,2\,\phi^A\phi^A\,\eta^{1\prime}\eta^{4\prime}
\nonumber\\
&&-\,4\,\bigl(\lambda\phi^A\pi^A
+\phi^{A\prime\prime}\pi^A\bigr)\,\eta^4\eta^4
+2\,\phi^A\pi^A\eta^{4\prime}\eta^{4\prime}
\nonumber\\
&&-\,4\,\phi^A\pi^A\,\eta^4\eta^{4\prime\prime}
-8\,\phi^{A\prime}\pi^A\eta^4\eta^{4\prime}
\nonumber\\
T_3^{(2)}&=&2\,\phi^A\phi^A\,\eta^1\eta^4
-\,2\,\phi^A\pi^A\eta^4\eta^4
\nonumber\\
T_4^{(2)}&=&4\,\bigl(\phi^A\pi^{A\prime}
+3\,\phi^{A\prime}\pi^A\bigr)\,\eta^1\eta^{4\prime}
+8\,\phi^A\pi^A\,\eta^1\eta^{4\prime\prime}
\label{4.18}
\end{eqnarray}

\bigskip\noindent
Contrary to the case without Lagrange multipliers, the quantities
which are obtained by summing $T_a$, $T_a^{(1)}$, and $T_a^{(2)}$ are
not first-class constraints. It would be necessary to go to higher
steps of the method. As one observes this is not an easy task,
because there might be other solutions in the second step, besides
the one we have found. We do not know which solution would be more
appropriate to use in the third step or, more than that, if this
solution would actually lead to a solution there.

\vspace{1cm}
\section{Conclusion}
\bigskip

We have considered the use of the BFFT formalism for systems with
nonlinear constraints. First, we have done a general analysis of the
conditions in which the method could be used without going to higher
order of the iterative process. In this way, we show why a particular
version of the $O(N)$ nonlinear sigma model can work for this case.
We also show that the same occurs for the supersymmetric case, but
leading to a nonlocal theory. We have also discussed the general case
and show that the full BFFT method is not feasible to be applied.  In
part because the first iterative step does not give a unique solution
and we do not know a priori what is the most convenient one to carry
out to the second step. More than that, we have also show that nor
all solutions of the first step lead to a solution in the second one.

\medskip
A way we envisage to circumvent the problems we have mentioned is to
consider the method in a less restrictive way, that is to say, by
considering that first-class constraints do not have to form just an
Abelian algebra, but to be open to the fact that a non-Abelian
algebra could be also possible~\cite{Barc}.

\vspace{1cm}
\noindent {\bf Acknowledgment:} I am in debt with R. Amorim and R.
Banerjee for useful discussion and comments. This work is supported
in part by Conselho Nacional de Desenvolvimento Cient\'{\i}fico e
Tecnol\'ogico - CNPq, Financiadora de Estudos e Projetos - FINEP and
Funda\c{c}\~ao Universit\'aria Jos\'e Bonif\'acio - FUJB (Brazilian
Research Agencies).

\vspace{1cm}
\appendix
\renewcommand{\theequation}{A.\arabic{equation}}
\setcounter{equation}{0}
\section*{Appendix}

\bigskip
One of the unexpected aspect of the BFFT method is that nor all
solution obtained in the first step of the method can lead to a
solution in the second step. In order to see this point with details,
let us consider another solution of eq. (\ref{4.9}), for example

\begin{eqnarray}
&&X_{11}=0\,,
\hspace{.5cm}X_{12}=0\,,
\hspace{.5cm}X_{13}=2\,\delta\,,
\hspace{.5cm}X_{14}=\phi^A\pi^A\,\delta\,,
\nonumber\\
&&X_{21}=\phi^A\phi^A\,\delta\,,
\hspace{.5cm}X_{22}=0\,,
\hspace{.5cm}X_{23}=0\,,
\nonumber\\
&&X_{24}=\frac{1}{2}\,\pi^A\pi^A\,\delta
-\frac{1}{2}\,\phi^A\phi^{A\prime\prime}\,\delta
-\frac{1}{2}\,\phi^A\phi^{A\prime}\,\delta^\prime\,,
\nonumber\\
&&X_{31}=0\,,
\hspace{.5cm}X_{32}=0\,,
\hspace{.5cm}X_{33}=0\,,
\hspace{.5cm}X_{34}=-\frac{1}{4}\,\phi^A\phi^A\,\delta\,,
\nonumber\\
&&X_{41}=0\hspace{.5cm}X_{42}=4\,\delta\,,
\hspace{.5cm}X_{43}=\delta^{\prime\prime}
+2\,\lambda\,\delta\,,
\hspace{.5cm}X_{44}=0\,.
\label{A.1}
\end{eqnarray}

\bigskip\noindent
We mention that $X_{23}$ and $X_{44}$ could have taken any value. The
choice above leads to the following equations in the second step of
the method

\begin{eqnarray}
&&\bigl\{2\eta^3+\phi^A\pi^A\eta^4,T_2^{(2)}\bigr\}_{(\eta)}
+\bigl\{T_1^{(2)},\phi^A\phi^A\eta^1-
\frac{1}{2}\,\phi^A\phi^{A\prime\prime}\eta^4
\nonumber\\
&&\phantom{\bigl\{2\eta^3+\phi^A\pi^A\eta^4}
-\frac{1}{2}\,\phi^A\phi^{A\prime}\eta^{4\prime}
+\frac{1}{2}\,\pi^A\pi^A\,\eta^4\bigr\}_{(\eta)}
=-\,2\,\phi^A\pi^A\eta^4\,\delta\,,
\label{A.2}\\
&&\bigl\{2\,\eta^3+\phi^A\pi^A\eta^4,T_3^{(2)}\bigr\}_{(\eta)}
-\,\frac{1}{4}\phi^A\phi^A\,
\bigl\{T_1^{(2)},\eta^4\bigr\}_{(\eta)}=0\,,
\label{A.3}\\
&&\bigl\{2\eta^3+\phi^A\pi^A\eta^4,T_4^{(2)}\bigr\}_{(\eta)}
+\bigl\{T_1^{(2)},4\,\eta^2+\eta^{3\prime\prime}
+2\lambda\eta^3\bigr\}_{(\eta)}
\nonumber\\
&&\phantom{\bigl\{2\eta^3+\phi^A\pi^A}
=-\,2\,\pi^A\pi^A\,\eta^4\,\delta
+2\,\lambda\,\phi^A\phi^A\,\eta^4\,\delta
+2\,\phi^A\phi^{A\prime\prime}\,\eta^4\,\delta
\nonumber\\
&&\phantom{\bigl\{2\eta^3+\phi^A\pi^A\eta^4}
+\,2\,\phi^A\phi^{A\prime}\,\eta^4\,\delta^\prime
+\phi^A\phi^A\eta^4\delta^{\prime\prime}\,,
\label{A.4}\\
&&\bigl\{\phi^A\phi^A\eta^1
-\frac{1}{2}\phi^A\phi^{A\prime\prime}\eta^4
-\frac{1}{2}\phi^A\phi^{A\prime}\eta^{4\prime}
+\frac{1}{2}\,\pi^A\pi^A\eta^4,T_3^{(2)}\bigr\}_{(\eta)}
\nonumber\\
&&\phantom{\bigl\{2\eta^3+\phi^A\pi^A\eta^4}
-\,\frac{1}{4}\phi^A\phi^A\,
\bigl\{T_2^{(2)},\eta^4\bigr\}_{(\eta)}
=-\,\frac{1}{2}\,\phi^A\phi^A\,\eta^4\,\delta\,,
\label{A.5}\\
&&\bigl\{\phi^A\phi^A\eta^1
-\frac{1}{2}\phi^A\phi^{A\prime\prime}\eta^4
-\frac{1}{2}\phi^A\phi^{A\prime}\eta^{4\prime}
+\frac{1}{2}\,\pi^A\pi^A\eta^4,T_4^{(2)}\bigr\}_{(\eta)}
\nonumber\\
&&\phantom{\bigl\{2\eta^3+\phi^A\pi^A\eta^4}
+\bigl\{T_2^{(2)},4\eta^2+\eta^{3\prime\prime}
+2\,\lambda\eta^3\bigr\}_{(\eta)}
=-\,4\,\phi^A\pi^A\eta^1\,\delta
\nonumber\\
&&\phantom{\bigl\{2\eta^3+\phi^A\pi^A\eta^4}
+\,3\,\phi^{A\prime\prime}\pi^A\eta^4\,\delta
+\,2\,\lambda\,\phi^A\pi^A\eta^4\,\delta
+\,\phi^{A\prime}\pi^A\eta^{4\prime}\,\delta
\nonumber\\
&&\phantom{\bigl\{2\eta^3+\phi^A\pi^A\eta^4}
+\phi^A\pi^{A\prime\prime}\eta^4\,\delta
+\phi^A\pi^{A\prime}\eta^{4\prime}\,\delta
+\,2\,\phi^A\pi^{A\prime}\eta^4\,\delta^\prime
\nonumber\\
&&\phantom{\bigl\{2\eta^3+\phi^A\pi^A\eta^4}
+\phi^A\pi^A\eta^{4\prime}\,\delta^\prime
+2\,\phi^{A\prime}\pi^A\eta^4\,\delta^\prime
+2\,\phi^A\pi^A\eta^4\,\delta^{\prime\prime}\,,
\label{A.6}\\
&&-\,\frac{1}{4}\,\phi^A\phi^A\,
\bigl\{\eta^4,T_4^{(2)}\bigr\}_{(\eta)}
+\bigl\{T_3^{(2)},4\eta^2+\eta^{3\prime\prime}
+2\,\lambda\eta^3\bigr\}_{(\eta)}
\nonumber\\
&&\phantom{\bigl\{2\eta^3+\phi^A\pi}
=\,\,2\,\eta^3\,\delta+\phi^A\pi^A\eta^4\,\delta
\label{A.7}
\end{eqnarray}

\bigskip
Of course, the best strategy to solve these equations is to start
from the most complicated one because in the simplest cases there are
much freedom in fixing the quantities $T_a^{(2)}$ and we do not know
a priori what would be the best choice we have to do. Let us then
consider eq.(\ref{A.6}). We notice that the right side can only be
matched by means of $T_2^{(2)}$.

\medskip
In order to fix the term with $\lambda$ in the right side of
(\ref{A.6}), we have two options. The first one is $T_2^{(2)}$ having
a term like

\begin{equation}
T_2^{(2)}=\phi^A\pi^A\eta^1\eta^4+\cdots
\label{A.8}
\end{equation}

\bigskip\noindent
where dots are representing other terms that we shall figure out
later.  With the choice above, we have

\begin{eqnarray}
\bigl\{T_2^{(2)},4\,\eta^2+\eta^{3\prime\prime}
+2\,\lambda\eta^3\bigr\}_{(\eta)}
&=&\bigl\{\phi^A\pi^A\,\eta^1\eta^4+\cdots,
4\,\eta^2+\eta^{3\prime\prime}
+2\,\lambda\eta^3\bigr\}_{(\eta)}
\nonumber\\
&=&2\,\lambda\,\phi^A\pi^A\,\eta^4\,\delta
-4\,\phi^A\pi^A\,\eta^1\,\delta
\nonumber\\
&&
+\phi^A\pi^A\,\eta^4\,\delta^{\prime\prime}+\cdots
\label{A.9}
\end{eqnarray}

\bigskip\noindent
We notice that the term $2\lambda\phi^A\pi^A\eta^4\delta$ was
actually obtained, but the term
$\phi^A\pi^A\eta^4\delta^{\prime\prime}$ does not match the
corresponding one in the right side of (\ref{A.6}). A solution could
be the introduction of a new term in the expression of $T_2^{(2)}$
like

\begin{equation}
T_2^{(2)}=\phi^A\pi^A\,\eta^1\eta^4
-\frac{1}{4}\,\phi^A\pi^A\,\eta^4\eta^{4\prime\prime}+\cdots
\label{A.10}
\end{equation}

\bigskip\noindent
In fact, this new term would give us the term that is lacking,
$\phi^A\pi^A\eta^4\delta^{\prime\prime}$. However, it will also lead
to the term $\phi^A\pi^A\eta^{4\prime\prime}\delta$, that does not
match any other term in (\ref{A.6}). We do not have any more options
to solve the problems we have.

\medskip
Let us recall what we have done till now. Looking at eq.(\ref{A.6}),
we had decided to fix the term with $\lambda$ in the right side, but
this did not work. Let us then forget the term in $\lambda$ and try
another solution, for example,

\begin{equation}
T_2^{(2)}=2\,\phi^A\pi^A\,\eta^1\eta^4+\cdots
\label{A.11}
\end{equation}

\bigskip\noindent
This choice yields

\begin{equation}
\bigl\{T_2^{(2)},4\,\eta+\eta^{3\prime\prime}
+2\,\lambda\,\eta^3\bigr\}_{(\eta)}
=-\,8\,\phi^A\pi^A\,\eta^1\delta
+2\,\phi^A\pi^A\,\eta^4\delta^{\prime\prime}
+4\,\lambda\,\phi^A\pi^A\,\eta^4\delta+\cdots
\label{A.12}
\end{equation}

\bigskip\noindent
There are two terms that do not match the right side of (\ref{A.6}).
We may then add another term in (\ref{A.11}), say

\begin{equation}
T_2^{(2)}=2\,\phi^A\pi^A\,\eta^1\eta^4
+\frac{1}{4}\,\lambda\,\phi^A\pi^A\,\eta^4\eta^4 +\cdots
\label{A.13}
\end{equation}

\bigskip\noindent
This leads to 

\begin{equation}
\bigl\{T_2^{(2)},4\,\eta+\eta^{3\prime\prime}
+2\,\lambda\,\eta^3\bigr\}_{(\eta)}
=-\,8\,\phi^A\pi^A\,\eta^1\delta
+2\,\phi^A\pi^A\,\eta^4\delta^{\prime\prime}
+2\,\lambda\,\phi^A\pi^A\,\eta^4\delta+\cdots
\label{A.14}
\end{equation}

\bigskip\noindent
However, the first term also does not match the corresponding one
into the right side of (\ref{A.6}). There is no other way we could
try in order to solve these problems. So, we cannot find a solution
for eq.(\ref{A.6})
\footnote{Even if we were able to obtain the so mentioned terms,
there would be more. For example, how could we match the terms
$\phi^{A\prime}\pi^A\,\eta^{4\prime}\delta$ and
$2\,\phi^{A\prime}\pi^A\,\eta^4\delta^\prime$? The factor 2 spoils
any attempt. The same occurs with
$2\,\phi^A\pi^{A\prime}\,\eta^4\delta^\prime$ and
$\phi^A\pi^{A\prime}\,\eta^{4\prime}\delta$.}.

\bigskip
We could then think to go back to the solution where we have fixed
(\ref{A.6}) and try to make some refinements on it. For example, we
had taken $X_{44}=0$, but we have also seen that it could have taken
any other value for it. Let us then conveniently make

\begin{equation}
X_{44}=\delta\,.
\label{A.15}
\end{equation}

\bigskip\noindent
Thus, the term $T_4^{(1)}$ turns to be

\begin{equation}
T_4^{(1)}=4\,\eta^2+\eta^{3\prime\prime}
+2\,\lambda\,\eta^3+\eta^4
\label{A.16}
\end{equation}

\bigskip\noindent
and consequently, instead of (\ref{A.6}) we have

\begin{eqnarray}
&&\bigl\{\phi^A\phi^A\,\eta^1
-\frac{1}{2}\,\phi^A\phi^{A\prime\prime}\,\eta^4
-\frac{1}{2}\phi^A\phi^{A\prime}\,\eta^{4\prime}
+\frac{1}{2}\pi^A\pi^A\,\eta^4,
T_4^{(2)}\bigr\}_{(\eta)}
\nonumber\\
&&\phantom{\bigl\{\phi^A\phi^A\,\eta^1}
+\bigl\{T_2^{(2)},4\,\eta^2+\eta^{3\prime\prime}
+2\,\lambda\,\eta^3+\eta^4\bigr\}_{(\eta)}
\nonumber\\
&&\phantom{\bigl\{\phi^A\phi^A}
=-\,4\,\phi^A\pi^A\,\eta^1\,\delta
+3\,\phi^{A\prime\prime}\pi^A\,\eta^4\,\delta
+2\,\lambda\,\phi^A\pi^A\,\eta^4\,\delta
\nonumber\\
&&\phantom{\bigl\{\phi^A\phi^A\,\eta^1}
+\phi^{A\prime}\pi^A\,\eta^{4\prime}\,\delta
+\phi^A\pi^{A\prime\prime}\,\eta^4\,\delta
+2\,\phi^A\pi^{A\prime}\,\eta^4\,\delta^\prime
\nonumber\\
&&\phantom{\bigl\{\phi^A\phi^A\,\eta^1}
+2\,\phi^A\pi^A\,\eta^4\,\delta^{\prime\prime}
+\phi^A\pi^{A\prime}\,\eta^{4\prime}\,\delta
+\phi^A\pi^A\,\eta^{4\prime}\,\delta^\prime
\nonumber\\
&&\phantom{\bigl\{\phi^A\phi^A\,\eta^1}
+2\,\phi^{A\prime}\pi^A\,\eta^1\,\delta^\prime
\label{A.17}
\end{eqnarray}

\bigskip\noindent
We now consider

\begin{equation}
T_2^{(2)}=\phi^A\pi^A\,\eta^1\eta^4
+\phi^A\pi^A\,\eta^{2\prime\prime}\eta^4
+\cdots
\label{A.18}
\end{equation}

\bigskip\noindent
and we then have

\begin{eqnarray}
&&\bigl\{T_2^{(2)},4\,\eta^2+\eta^{3\prime\prime}
+2\,\lambda\,\eta^3+\eta^4\big\}_{(\eta)}
=-\,4\,\phi^A\pi^A\,\eta^1\,\delta
\nonumber\\
&&\phantom{\bigl\{T_2^{(2)},4\,\eta^2}
+\,2\,\phi^A\pi^A\,\eta^4\,\delta^{\prime\prime}
+2\,\lambda\,\phi^A\pi^A\,\eta^4\,\delta
-4\,\phi^A\pi^A\,\eta^{2\prime\prime}\,\delta
+\cdots
\label{A.19}
\end{eqnarray}

\bigskip\noindent
We notice that the term $2\,\phi^A\pi^A\eta^4\delta^{\prime\prime}$
was obtained, but the price paid was also the obtainment of the term
$-\,4\,\phi^A\pi^A\eta^{2\prime\prime}\delta$ that does not match any
term in the right side of (\ref{A.6}). There is no other choice for
$X_{44}$ that could be more convenient. Definitely, eq. (\ref{A.6})
does not have solution. 

\bigskip
It is opportune to mention that there are many other attempts that do
not work also. Let us write down some of these examples,

\begin{eqnarray}
&&X_{11}=-\,2\,\phi^A\phi^A\delta\,,
\hspace{.5cm}X_{12}=0\,,
\hspace{.5cm}X_{13}=0\,,
\hspace{.5cm}X_{14}=4\,\phi^A\pi^A\delta\,,
\nonumber\\
&&X_{21}=0\,,
\hspace{.5cm}X_{22}=0\,,
\hspace{.5cm}X_{23}=\delta\,,
\hspace{.5cm}X_{24}=2\,\pi^A\pi^A\delta\,,
\nonumber\\
&&X_{31}=0\,,
\hspace{.5cm}X_{32}=0\,,
\hspace{.5cm}X_{33}=0\,,
\hspace{.5cm}X_{34}=-\,\phi^A\phi^A\delta\,,
\nonumber\\
&&X_{41}=-\,2\,\lambda\,\phi^A\phi^A\,\delta
-2\,\phi^A\phi^{A\prime\prime}\,\delta
-2\,\phi^A\phi^{A\prime}\delta^\prime
-\phi^A\phi^A\delta^{\prime\prime}\,,
\nonumber\\
&&X_{42}=\delta\,,
\hspace{.5cm}X_{43}=0\,,
\hspace{.5cm}X_{44}=0\,.
\label{A.20}
\end{eqnarray}

\bigskip

\begin{eqnarray}
&&X_{11}=-\,2\,\phi^A\phi^A\delta\,,
\hspace{.5cm}X_{12}=0\,,
\hspace{.5cm}X_{13}=0\,,
\hspace{.5cm}X_{14}=4\,\phi^A\pi^A\delta\,,
\nonumber\\
&&X_{21}=0\,,
\hspace{.5cm}X_{22}=0\,,
\hspace{.5cm}X_{23}=\delta\,,
\nonumber\\
&&X_{24}=-\,2\,\lambda\,\phi^A\phi^A\,\delta
-2\,\phi^A\phi^{A\prime\prime}\,\delta
-2\,\phi^A\phi^{A\prime}\delta^\prime
-\phi^A\phi^A\delta^{\prime\prime}\,,
\nonumber\\
&&X_{31}=0\,,
\hspace{.5cm}X_{32}=0\,,
\hspace{.5cm}X_{33}=0\,,
\hspace{.5cm}X_{34}=-\,\phi^A\phi^A\delta\,,
\nonumber\\
&&X_{41}=2\,\pi^A\pi^A\,\delta\,,
\hspace{.5cm}X_{42}=\delta\,,
\hspace{.5cm}X_{43}=0\,,
\hspace{.5cm}X_{44}=0\,.
\label{A.21}
\end{eqnarray}

\bigskip
\begin{eqnarray}
&&X_{11}=0\,,
\hspace{.5cm}X_{12}=0\,,
\hspace{.5cm}X_{13}=0\,,
\hspace{.5cm}X_{14}=\delta\,,
\nonumber\\
&&X_{21}=0\,,
\hspace{.5cm}X_{22}=2\,\phi^A\phi^A\delta\,,
\hspace{.5cm}X_{24}=0\,,
\nonumber\\
&&X_{23}=2\,\pi^A\pi^A\,\delta
-2\,\lambda\phi^A\phi^A\,\delta
-2\,\phi^A\phi^{A\prime\prime}\,\delta
-2\,\phi^A\phi^{A\prime}\delta^\prime
-\phi^A\phi^A\delta^{\prime\prime}\,,
\nonumber\\
&&X_{31}=0\,,
\hspace{.5cm}X_{32}=0\,,
\hspace{.5cm}X_{33}=-\,\phi^A\phi^A\delta\,,
\hspace{.5cm}X_{34}=0\,,
\nonumber\\
&&X_{41}=\delta\,,
\hspace{.5cm}X_{42}=4\,\phi^A\pi^A\delta\,,
\hspace{.5cm}X_{43}=0\,,
\hspace{.5cm}X_{44}=0\,.
\label{A.22}
\end{eqnarray}

\bigskip
\begin{eqnarray}
&&X_{11}=0\,,
\hspace{.5cm}X_{12}=0\,,
\hspace{.5cm}X_{13}=0\,,
\hspace{.5cm}X_{14}=\delta\,,
\nonumber\\
&&X_{21}=-\,2\,\pi^A\pi^A\,\delta
+2\,\phi^A\phi^{A\prime\prime}\,\delta\,
+2\,\phi^A\phi^{A\prime}\delta^\prime\,,
\nonumber\\
&&X_{22}=2\,\phi^A\phi^A\delta\,,
\hspace{.5cm}X_{23}=0\,,
\hspace{.5cm}X_{24}=0\,,
\nonumber\\
&&X_{31}=\phi^A\phi^A\delta\,,
\hspace{.5cm}X_{32}=0\,,
\hspace{.5cm}X_{33}=0\,,
\hspace{.5cm}X_{34}=0\,,
\nonumber\\
&&X_{41}=0\,,
\hspace{.5cm}X_{42}=4\,\phi^A\pi^A\delta\,,
\hspace{.5cm}X_{43}=\delta\,,
\hspace{.5cm}X_{44}=\lambda\,\delta
+\frac{1}{2}\,\delta^{\prime\prime}\,.
\label{A.23}
\end{eqnarray}

\bigskip\noindent
There are many other solutions, but just corresponding to
permutations among the coefficients of the solutions above. It is
important to mention that (\ref{A.20}) corresponds to the case
discussed in the text for a different value of the coefficient
$X_{44}$. The fact that suggest me to concentrate in this solution
and not on the others is that it was the first case where we could
find a solution for the most complicated of the equations. In all
other cases, this did not occur.

\end{document}